\documentclass[11pt,twoside]{article}
\usepackage{graphicx}
\usepackage{amsmath}
\usepackage{amssymb}
\usepackage{mathrsfs}

\begin{document}

\thispagestyle{plain}

\newcommand{\pst}{\hspace*{1.5em}}
\newcommand{\be}{\begin{equation}}
\newcommand{\ee}{\end{equation}}
\newcommand{\ds}{\displaystyle}
\newcommand{\bdm}{\begin{displaymath}}
\newcommand{\edm}{\end{displaymath}}
\newcommand{\beq}{\begin{eqnarray}}
\newcommand{\eeq}{\end{eqnarray}}

\begin{center} {\Large \bf
\begin{tabular}{c}
Operators and their symbols in the optical probabilistic \\[-1mm]
representation of quantum mechanics
\end{tabular}
 } \end{center}

\smallskip

\begin{center} {\bf G. G. Amosov$^1$, Ya. A. Korennoy$^2$, V. I. Man'ko$^2$ }\end{center}

\smallskip

\begin{center}
{\it $^1$
Steklov Mathematical Institute,  \\
ul. Gubkina 8, Moscow 119991, Russia}\\
{\it $^2$P.N.    Lebedev Physics Institute,                          \\
       Leninsky prospect 53, 117924, Moscow, Russia }
\end{center}

\begin{abstract}\noindent
Explicit expressions for most interesting quantum operators
in optical tomography representation are found.
General formalism of symbols of operators is presented in optical 
tomographic representation.
The symbols of the operators are found explicitly for physical quantities.
\end{abstract}


\section{Introduction}

The transition from quantum to classical mechanics has been an important research subject
since the beginning of quantum mechanics (see \cite{VV1, VV2} for a review). A suitable setting for this
problem is represented by the Wigner-Weyl-Moyal formalism where the operators corresponding
to observables and the states, considered as linear functionals on the space of observables, are
mapped onto functions on a suitable manifold. Such a representation for quantum mechanics
has been later generalized yielding to the deformation quantization program \cite{VV3}. There the
operator noncommutativity is implemented by a noncommutative (star) product which is a
generalization of the Moyal product \cite{VV4,VV5,VV6}.
Since then, most attention to the star-product
quantization scheme has been devoted to the case where the functions (symbols of the operators)
are defined on the “classical” phase space of the system \cite{VV7,VV8,VV9,VV10}.

In a different setting it was recently established \cite{V26, V27} that the symplectic \cite{V28, V29,V29_1} and
spin \cite{V30, V31} tomography, which furnish alternative formulations of quantum mechanics
and quantum field theory \cite{V33}, can be described as well within a star-product scheme.
Moreover, in \cite{V27} different known star-product schemes were presented in a unified form.
There, the symbols of the operators are defined in terms of a special family of operators
using the trace formula (what we sometimes call the ‘dequantization’ map because of its
original meaning in the Wigner-Weyl formalism), while the reconstruction of operators
in terms of their symbols (the ‘quantization’ map) is determined using another family of
operators. These two families determine completely the star-product scheme, including
the kernel of the star-product.

The aim of our work is to find the explicit expressions of most physically interesting 
operators and their dual symbols in optical tomographic representation, nesessary for practical
calculations.

The paper is organized as follows. In Sec.2 we review optical tomography of quantum states.
In Sec.3 the correspondence rules for physical operators in optical tomographic
representation are found.
In Sec.4 the general formalism of symbols of operators is presented in optical 
tomographic representation. 
The expressions for the dual symbols of the operators
in terms of singular generalized functions
 and for the kernel of their star-product
are presented. 
In Sec.5 the representation of dual symbols of operators in terms of regular
generalized functions is given.

\section{Optical tomographic representation of quantum states}
In this section we give a short review of the tomographic representation of
quantum mechanics by using so called optical tomogram \cite{berber,vogrisk}.
For the photon states this tomogram is measured experimentally 
\cite{raymer,Lvov. Ray Rev.Mod.Phys}.
For simplicity of the formulas we consider a case of one degree of
freedom with dimensionless variables,
because the generalization of all our calculations and results
to the case of any arbitrary number dimensional
degrees of freedom is obvious.

If we have the density matrix of the quantum state $\hat\rho$, the optical 
tomogram is defined as
\begin{equation}		\label{eq1}
w(X,\theta)= \mbox{Tr}\{\hat\rho\delta(X-\hat q\cos\theta-\hat p\sin\theta)\}
=\langle X,\theta\vert\hat\rho\vert X,\theta\rangle,
\end{equation}
where $\vert X,\theta\rangle$ is an eigenvector of the hermitian
operator $\hat q\cos\theta+\hat p\sin\theta$ for the eigenvalue $X$
\begin{equation}		\label{eq2}
\langle q\vert X,\theta\rangle=\frac{1}{\sqrt{2\pi\vert\sin\theta\vert}}
\exp\left(i\frac{Xq-\frac{q^2}{2}\cos\theta}{\sin\theta}\right).
\end{equation}

In terms of the Wigner function \cite{wig32} the tomogram
$w(X,\theta)$ is expressed as
\begin{equation}		\label{eq4}
w(X,\theta)=\frac{1}{(2\pi)^2}\int W(q,p)e^{i\eta(X-q\cos\theta-
p\sin\theta)}{\mbox d}\eta~{\mbox d}q~{\mbox d}p.
\end{equation}
This relation can be reversed using the symmetry property
 of the optical tomogram
\be		\label{eq4_1}
w(X,\theta,t)=w((-1)^k X,\theta+\pi k,t),~~~~k=0,~\pm1,~\pm2, ...
\ee
After calculations we can write
\begin{equation}		\label{eq5}
W(q,p)=\frac{1}{2\pi}\int\limits_{0}^{\pi}~\mbox{d}\theta
\int\limits_{-\infty}^{+\infty}\int\limits_{-\infty}^{+\infty}w(X,\theta)
|\eta |~e^{i\eta(X-q\cos\theta-p\sin\theta)} \mbox{d}\eta~\mbox{d}X.
\end{equation}
From (\ref{eq4}) using the relations between the Wigner function and the density
matrix $\rho(q,q')$ in coordinate representation
\begin{equation}		\label{eq6}
W(q,p)=\int\rho(q+u/2,q-u/2)e^{-ipu}{\mbox d}u,
\end{equation}
\begin{equation}		\label{eq7}
\rho(q,q')=\frac{1}{2\pi}\int W\left(\frac{q+q'}{2},p\right)
e^{ip(q-q')}{\mbox d}p,
\end{equation}
we can write the relations between the optical tomogram and the density
matrix in coordinate representation as follows
\be		\label{eq8}
w(X,\theta)=\frac{1}{2\pi}\int\rho\left(
q+\frac{u\sin\theta}{2},q-\frac{u\sin\theta}{2}
\right)e^{-iu(X-q\cos\theta)}{\mbox d}u~{\mbox d}q,
\ee
\be		\label{eq9}
\rho(q,q')=\frac{1}{2\pi}\int\limits_{0}^{\pi}~\mbox{d}\theta
\int\limits_{-\infty}^{+\infty} w(X,\theta)\vert\eta\vert
\exp\left\{i\eta\left(X-\frac{q+q'}{2}\cos\theta\right)\right\}
\delta(q-q'-\eta\sin\theta)~{\mbox d}\theta~{\mbox d}\eta~{\mbox d}X.
\ee
Thus, the tomogram $w(X,\theta)$ contains all information about
the quantum state.

The evolution equation and energy level equation for optical tomograms
were found explicitly in \cite{Kor13}.

\section{The correspondence rules for physical operators in optical tomographic representation}
Using the relation (\ref{eq7}) between the density matrix and the Wigner function 
one has the correspondence rules
\be		\label{eq_25}
\frac{\partial\rho}{\partial t} \longleftrightarrow\frac{\partial W}{\partial t},~~~~
\frac{\partial\rho}{\partial x} \longleftrightarrow\left(\frac{1}{2}\frac{\partial}{\partial q}+ip\right)W,
\ee
\bdm
\frac{\partial\rho}{\partial x'} \longleftrightarrow\left(\frac{1}{2}\frac{\partial}{\partial q}-ip\right)W;~~~~
x\rho \longleftrightarrow\left(q+\frac{i}{2}\frac{\partial}{\partial p}\right)W;~~~~
x'\rho \longleftrightarrow\left(q-\frac{i}{2}\frac{\partial}{\partial p}\right)W.
\edm

Using the relation  (\ref{eq4})  between the  Wigner function and the optical tomogram 
one can find the correspondence rules for the operators acting on the Wigner function 
and the optical tomogram:
\be		\label{eq_27}
\cos\theta\frac{\partial}{\partial X}w(X,\theta)=
\frac{1}{4\pi^2}\int ik\cos\theta~e^{-ikq\cos\theta}
W(q,p,t)~e^{ik(X-p\sin\theta)}\mbox{d}k~\mbox{d}q~\mbox{d}p.
\ee
In view of the equality
\bdm
ik\cos\theta~e^{-ikq\cos\theta}=-\frac{\partial}{\partial q}~e^{-ikq\cos\theta}
\edm
and integrating (\ref{eq_27}) by parts with account of 
$W(q,p)\to0$ for $q\to\pm \infty $ we arrive at 
\be		\label{eq_28}
\cos\theta\frac{\partial}{\partial X}w(X,\theta)=
\frac{1}{4\pi^2}\int \frac{\partial W(q,p)}{\partial q}\delta(X-q\cos\theta-p\sin\theta)\mbox{d}q~\mbox{d}p.
\ee
It means that
\be		\label{eq_29}
\frac{\partial}{\partial q}W(q,p) \longleftrightarrow\cos\theta\frac{\partial}{\partial X}w(X,\theta).
\ee
Analogously we have the correspondence rule
\be		\label{eq_30}
\frac{\partial}{\partial p}W(q,p) \longleftrightarrow\sin\theta\frac{\partial}{\partial X}w(X,\theta).
\ee
Applying the operator $\left(\partial /\partial X\right)^{-1}$ which is defined by acting on plane wave as 
\be		\label{eq_31}
\left(\frac{\partial}{\partial X}\right)^{-1}~e^{ikX}=\frac{1}{ik}~e^{ikX}
\ee
to (\ref{eq4}) together with differentiation over the angle variable $\theta$ and multiplication by $\sin\theta$ we get the equality
\be		\label{eq_32}
\sin\theta\frac{\partial}{\partial \theta}\left(\frac{\partial}{\partial X}\right)^{-1}w(X,\theta)
+X\cos\theta~w(X,\theta)
=\frac{1}{4\pi^2}\int q~W(q,p)~e^{ik(X-q\cos\theta-p\sin\theta)}\mbox{d}k~\mbox{d}q~\mbox{d}p.
\ee
We used identities
\bdm
(q\sin\theta-p\cos\theta)\sin\theta+X\cos\theta=q+(X-q\cos\theta-p\sin\theta)\cos\theta,
\edm
\bdm
\delta(X-q\cos\theta-p\sin\theta)~(X-q\cos\theta-p\sin\theta)=0.
\edm
The relation (\ref{eq_32}) gives the correspondence rule
\be		\label{eq_33}
q~W(q,p) \longleftrightarrow \left(\sin\theta\left(\frac{\partial}{\partial X}\right)^{-1}\frac{\partial}{\partial\theta}
+X\cos\theta\right)w(x,\theta).
\ee
Analogously we get the correspondence rule
\be		\label{eq_34}
p~W(q,p) \longleftrightarrow \left(-\cos\theta\left(\frac{\partial}{\partial X}\right)^{-1}\frac{\partial}{\partial\theta}
+X\sin\theta\right)w(x,\theta).
\ee

If the product of the direct and the inverse Radon transform gives the unity,
then the explicit form in the optical tomographic representation
of the product  of the operators  is equal to the product of these operators
in the optical tomographic representation.

More over, suppose  we have a set of operators $\{\hat{\cal A}_{ik}\},$
acting on the set of functions $W({\vec q},{\vec p})\in{\cal S}^{2n}$ (we consider a multidimentional case), 
and let for any $W({\vec q},{\vec p})\in{\cal S}^{2n},$ we have $\hat{\cal A}_{ik}~W({\vec q},{\vec p})\in{\cal S}^{2n}$
for any $\hat{\cal A}_{ik}~\in~\{\hat{\cal A}_{ik}\},$ then we can write
\be                             \label{r38_1}
{\bf\mbox R}\left[\sum_iC_i\prod_k(\hat{\cal A}_{ik})^{l_k}
W({\vec q},{\vec p})\right]({\vec X},{\vec\theta})=
\sum_iC_i\prod_k\left(
{\bf\mbox R}\left[\hat{\cal A}_{ik}^{l_k}\right]({\vec X},{\vec\theta})\right)^{l_k}
{\bf\mbox R}\left[W({\vec q},{\vec p})\right]({\vec X},{\vec\theta}),
\ee
where $\sum_i$ and $\prod_k$ -- no more than countable. 

Using the formulas given in this paragraph, it is possible to find explicit form
of any  interesting for the practice operators
in optical probability representation.

Thus one can find the operators $\hat{\vec q},$ $\hat{\vec p},$
$\hat{\vec q}^2,$ $\hat{\vec p}^2,$ $\hat{\vec q}\hat{\vec p}$
in the optical tomographic representation
\begin{eqnarray}
\hat q_i&=&\sin\theta_i\left[\frac{\partial}{\partial X_i}\right]^{-1}
\frac{\partial}{\partial\theta_i}+X_i\cos\theta_i+\frac{i}{2}
\frac{\hbar}{m_i\omega_{oi}}\sin\theta_i\frac{\partial}{\partial X_i};
\nonumber \\[3mm]
\hat p_i&=&m_i\omega_{0i}\left(-\cos\theta_i\left[\frac{\partial}{\partial X_i}\right]^{-1}
\frac{\partial}{\partial\theta_i}+X_i\sin\theta_i\right)-\frac{i\hbar}{2}
\cos\theta_i\frac{\partial}{\partial X_i};
\nonumber \\[3mm]
\hat q_i^2&=&\sin^2\theta_i\left[\frac{\partial}{\partial X_i}\right]^{-2}
\left(\frac{\partial^2}{\partial\theta_i^2}+1\right)
+X_i\left[\frac{\partial}{\partial x_i}\right]^{-1}
\left(
\sin2\theta_i\frac{\partial}{\partial\theta_i}-\sin^2\theta_i\right)
+X_i^2\cos^2\theta_i\nonumber \\[3mm]
&+&i\frac{\hbar}{m_i\omega_{oi}}\left\{
\sin^2\theta_i\frac{\partial}{\partial\theta_i}+
\frac{\sin2\theta_i}{2}\left(1+X_i\frac{\partial}{\partial x_i}\right)\right\}
-\frac{1}{4}\frac{\hbar^2}{m_i^2\omega_{oi}^2}\sin^2\theta_i
\frac{\partial^2}{\partial X^2}; \nonumber \\[3mm]
\hat p_i^2&=&m_i^2\omega_{oi}^2\left\{
\cos^2\theta_i\left[\frac{\partial}{\partial x_i}\right]^{-2}
\left(\frac{\partial^2}{\partial\theta_i^2}+1\right)
-X_i\left[\frac{\partial}{\partial x_i}\right]^{-1}
\left(
\sin2\theta_i\frac{\partial}{\partial\theta_i}+\cos^2\theta_i\right)
+X_i^2\sin^2\theta_i\right\} \nonumber \\[3mm]
&-&i\hbar m_i\omega_{0i}\left\{
\cos^2\theta_i\frac{\partial}{\partial\theta_i}-
\frac{\sin2\theta_i}{2}\left(1+X_i\frac{\partial}{\partial X_i}\right)\right\}
-\frac{\hbar^2}{4}\cos^2\theta_i
\frac{\partial^2}{\partial X^2}; \nonumber \\[3mm]
\hat q_i\hat p_i&=&m_i^2\omega_{oi}^2\left\{
-\frac{\sin2\theta_i}{2}\left[\frac{\partial}{\partial X_i}\right]^{-2}
\left(\frac{\partial^2}{\partial\theta_i^2}+1\right)
+X_i\left[\frac{\partial}{\partial X_i}\right]^{-1}
\left(
\frac{\sin2\theta_i}{2}-\cos2\theta_i\frac{\partial}{\partial\theta_i}\right)
+X_i^2\frac{\sin^2\theta_i}{2}\right\} \nonumber \\[3mm]
&-&i\hbar\left\{
\frac{X_i}{2}\frac{\partial}{\partial X_i}\cos^2\theta_i+
\frac{\sin2\theta_i}{2}\frac{\partial}{\partial\theta_i}-
\sin^2\theta_i\right\}
+\frac{\hbar^2}{8m_i\omega_{oi}}\sin^2\theta_i
\frac{\partial^2}{\partial X^2}.
\label{r46_1}
\end{eqnarray}

Let us find the momentum operator in the optical tomographic representation.
As known in the density matrix representation 
$\hat l=-i\hbar[\vec q,\nabla_{\vec q}],$ i.e. 
$\hat l_1=\hat q_2\hat p_3-\hat p_2\hat q_3,$ $l_2$ and $l_3$ are given from the relation for $\hat l_1$ by 
cyclic replacement of indices. In the Wigner representation
\bdm
\hat l_1=-i\left\{\frac{q_2}{2}\frac{\partial}{\partial q_3}+
iq_2p_3+\frac{i}{4}\frac{\partial^2}{\partial q_3\partial p_2}
-\frac{p_3}{2}\frac{\partial}{\partial p_2}-\frac{q_3}{2}
\frac{\partial}{\partial q_2}-iq_3p_2-\frac{i}{4}
\frac{\partial^2}{\partial p_3\partial q_2}
+\frac{p_2}{2}\frac{\partial}{\partial p_3}\right\},
\edm
and corresponding Wigner symbol of this operator
\bdm
W_{\hat l_1}(\vec  q,\vec p)=q_2p_3-q_3p_2.
\edm
In the optical distribution representation 
\begin{eqnarray}
\hat l_1&=&-i\left\{\frac{1}{2}\left(\sin\theta_2
\left[\frac{\partial}{\partial X_2}\right]^{-1}
\frac{\partial}{\partial\theta_2}+X_2\cos\theta_2\right)
\cos\theta_3\frac{\partial}{\partial X_3}\right. \nonumber \\[3mm]
&+&
i\left(\sin\theta_2\left[\frac{\partial}{\partial X_2}\right]^{-1}
\frac{\partial}{\partial\theta_2}+X_2\cos\theta_2\right)
\left(-\cos\theta_3\left[\frac{\partial}{\partial X_3}\right]^{-1}
\frac{\partial}{\partial\theta_3}+X_3\sin\theta_3\right) \nonumber \\[3mm]
&+&\left.\frac{i}{4}\sin\theta_2\frac{\partial}{\partial X_2}
\cos\theta_3\frac{\partial}{\partial X_3} 
+
\frac{\sin\theta_2}{2}\frac{\partial}{\partial X_2}
\left(\cos\theta_3\left[\frac{\partial}{\partial X_3}\right]^{-1}
\frac{\partial}{\partial\theta_3}-X_3\sin\theta_3\right)\right\}
+i\left\{\frac{}{}2\leftrightarrow3\right\}.\nonumber \\
\label{r46_5}
\end{eqnarray}
Components $\hat l_2$ and $\hat l_3$ are given from  (\ref{r46_5}) by cyclic
replacement of indices.

The creation and annihilation operators acting on the density matrix
in coordinate representation have the form
\be
\hat a=\frac{1}{\sqrt2}\left(q+\frac{\partial}{\partial q}\right);~~~~
\hat a^{\dag}=\frac{1}{\sqrt2}\left(q-\frac{\partial}{\partial q}\right) 
\ee

In the optical probability representation

\begin{eqnarray}
\hat a_i&=&\frac{\exp(i\theta_i)}{\sqrt2}\left\{\frac{1}{2}\frac{\partial}{\partial X_i}
+X_i-i\left[\frac{\partial}{\partial X_i}\right]^{-1}
\frac{\partial}{\partial\theta_i}\right\},\nonumber \\[3mm]
\hat a^{\dag}_i&=&\frac{\exp(-i\theta_i)}{\sqrt2}\left\{\frac{1}{2}\frac{\partial}{\partial X_i}
+X_i+i\left[\frac{\partial}{\partial x_i}\right]^{-1}
\frac{\partial}{\partial\theta_i}\right\}.
\label{r34_4}
\end{eqnarray}

For the number of quanta operator 
$\hat N_i=\hat a_i^{\dag}\hat a_i$ in $i-$th mode of $n-$dimentional oscillator 
we have
\bdm
\hat N_i\rho(\vec q,\vec q')=\hat a_i^{\dag}a_i\rho(\vec q,\vec q')=
\frac{1}{2}\left\{q_i^2-\frac{\partial^2}{\partial q_i^2}
-1\right\}\rho(\vec q,\vec q'),
\edm
in the Wigner representation
\begin{eqnarray}
(\hat N_i)_WW(\vec q,\vec p)&=&(\hat a_i^{\dag})_W(\hat a_i)_W
W(\vec q,\vec p)\nonumber \\[3mm]
&=&\frac{1}{2}\left\{q_i^2-\frac{1}{4}
\left(\frac{\partial^2}{\partial p_i^2}+\frac{\partial^2}{\partial q_i^2}
\right)+iq_i\frac{\partial}{\partial p_i}-ip_i\frac{\partial}{\partial q_i}+
p_i^2-1\right\}W(\vec q,\vec p).\nonumber
\end{eqnarray}
Using the formulas  of Sec.3 of this work, or taking the product of two operators (\ref{r34_4}) we arrive at
\begin{eqnarray}                             
\hat N_iw(\vec X,\vec\theta)&=&\hat a_i^{\dag}\hat a_i
w(\vec X,\vec\theta)\nonumber \\[3mm]
&=&\frac{1}{2}\left\{
\left[\frac{\partial}{\partial X_i}\right]^{-2}
\left(\frac{\partial^2}{\partial\theta_i^2}+1\right)
+X_i^2
-X_i\left[\frac{\partial}{\partial X_i}\right]^{-1}
-\frac{1}{4}\frac{\partial^2}{\partial X_i^2}
+i\frac{\partial}{\partial\theta_i}-1\right\}
w(\vec X,\vec\theta).\nonumber \\
\label{r38_3}
\end{eqnarray}
The operator $\hat N_i$ acts on the functions $w_n(\vec X,\vec\theta)$
of harmonic oscillator according to the formula
\be                             \label{r38_4}
\hat N_iw_{n_i}(\vec X,\vec\theta)=n_iw_{n_i}(\vec X,\vec\theta),
\ee
where  $n_i$  is a number of quanta  in the $i-$th mode.
The result  (\ref{r38_3}) also can be found as the product of creation
and annihilation operators in the optical probability representation.

Note, that at a derivation  of correspondence rules we actually used, 
that functions $W(\vec q,\vec p)$ belong to 
the space \cite{Gelfand}
of  well-behaved test functions ${\cal S}^{2n},$ on which the space of the
generalized functions of slow growth ${\cal S}'^{2n}$ can be constructed.

\section{General formalism of symbols of operators}
The relation between the density matrix and the tomogram can be
represented in the form
\bdm
w(X,\theta)=\mbox{Tr}\{\hat \rho\hat U(X,\theta)\},~~~
\hat\rho=\int w(X,\theta)\hat D(X,\theta) \mbox{d}X~\mbox{d}\theta,
\edm
where
\bdm
\hat U(X,\theta)=\delta(X\hat 1-\hat q\cos\theta-\hat p\sin\theta),
\edm
\bdm
\hat D(X,\theta)=\frac{1}{2\pi}\int\vert\eta\vert
~e^{\ds{i\eta(X-\hat q\cos\theta-\hat p\sin\theta)}}
\mbox{d}\eta
\edm
are the dequantizer and quantizer operator respectively.
These operators satisfy to orthogonality and completeness conditions.
\be		\label{ortog}
\mbox{Tr}\{\hat U(X,\theta)\hat D(X',\theta')\}=
\delta(X\cos(\theta-\theta')-X')\delta(\sin(\theta-\theta')),
\ee
\be		\label{compl}
\int\hat D_{\hat q'\hat p'}(X,\theta)
\hat U_{\hat q\hat p}(X,\theta)\,\mbox{d}X\,\mbox{d}\theta=
\delta(\hat q-\hat q')\delta(\hat p-\hat p').
\ee
Let us associate the symbol $w_{\hat A}(X,\theta)$ to the arbitrary operator $\hat A$
by the definition
\bdm
w_{\hat A}(X,\theta)=\mbox{Tr}\{\hat A\hat U(X,\theta)\}.
\edm
Taking into account the completeness condition (\ref{compl}) we can write 
the inverse relation
\bdm
\hat A=\int w_{\hat A}(X,\theta)\hat D(X,\theta) \mbox{d}X~\mbox{d}\theta.
\edm
The action of the operator $\hat A$ to the density matrix can be represented
in tomographic representation as the integral operator
\bdm
\mbox{Tr}\{\hat A\hat \rho\hat U(X,\theta) \}=
\int w_{\hat A}(X',\theta')w(X'',\theta'')
\mbox{Tr}\{ \hat D(X',\theta')\hat D(X'',\theta'')\hat U(X,\theta)\}
\mbox{d}X'~\mbox{d}\theta'~\mbox{d}X''~\mbox{d}\theta''.
\edm

The average value of the operator $\hat A$ equals
\bdm
\mbox{Tr}\{\hat A\hat \rho \}=
\int w(X,\theta)\mbox{Tr}\{\hat A\hat D(X,\theta) \}
\mbox{d}X~\mbox{d}\theta=
\int w(X,\theta)w_{\hat A}^{(d)}(X,\theta)
\mbox{d}X~\mbox{d}\theta,
\edm
where we denote the designation for dual symbol of the
operator $\hat A$ 
\be			\label{dual}
w_{\hat A}^{(d)}(X,\theta)=\mbox{Tr}\{\hat A\hat D(X,\theta) \}.
\ee
With the help of (\ref{compl}) the operator $\hat A$ can be found from
its dual symbol  
\bdm
\hat A=\int w_{\hat A}^{(d)}(X,\theta)\hat U(X,\theta) \mbox{d}X~\mbox{d}\theta.
\edm
Symbol $w_{\hat A}(X,\theta)$ and corresponding dual symbol
$w_{\hat A}^{(d)}(X,\theta)$ are associated by the relations
\bdm
w_{\hat A}^{(d)}(X,\theta)=
\int w_{\hat A}(X',\theta')
\mbox{Tr}\{ \hat D(X',\theta')\hat D(X,\theta)\}
\mbox{d}X'~\mbox{d}\theta',
\edm
\bdm
w_{\hat A}(X,\theta)=
\int w_{\hat A}^{(d)}(X',\theta')
\mbox{Tr}\{ \hat U(X',\theta')\hat U(X,\theta)\}
\mbox{d}X'~\mbox{d}\theta'.
\edm

The dual symbol of the product of two operators $\hat A$ and $\hat B$
is equal to the star-product with the corresponding kernel
\begin{eqnarray}		
w_{\hat A\hat B}^{(d)}(X,\theta)&=&
w_{\hat A}^{(d)}(X,\theta)*w_{\hat B}^{(d)}(X,\theta) \nonumber \\[3mm]
&=&\int K^{(d)}(X,\theta;X',\theta';X'',\theta'')
w_{\hat A}^{(d)}(X',\theta')w_{\hat B}^{(d)}(X'',\theta'')
\mbox{d}X'~\mbox{d}\theta'~\mbox{d}X''~\mbox{d}\theta'',
\label{starDual}
\end{eqnarray}
where
\be		\label{kernelD}
K^{(d)}(X,\theta;X',\theta';X'',\theta'')=
\mbox{Tr}\{ \hat U(X',\theta')\hat U(X'',\theta'')\hat D(X,\theta)\}.
\ee
This formula can be transformed to a form suitable for practical use as follows
\begin{eqnarray}		
K^{(d)}(X,\theta;X',\theta';X'',\theta'')&=&
\frac{1}{(2\pi)^2}\int \delta(X'-q\cos\theta'-p\sin\theta')
\delta(X''-q\cos\theta''-p\sin\theta'')\vert\eta\vert \nonumber \\[3mm]
&\times&\exp\{i\eta(X-q\cos\theta-p\sin\theta)\}
\exp\left\{ i\eta^2\frac{\sin(\theta-\theta')\sin(\theta-\theta'')}{\sin(\theta'-\theta'')}\right\}
\mbox{d}\eta~\mbox{d}q~\mbox{d}p. \nonumber \\
\label{kernelD1}
\end{eqnarray}

From the definition of dual symbol (\ref{dual}) for the operators $\hat 1$,
$\hat q$ and $\hat p$ after calculations we arrive at the formulas
\bdm
w_{\hat 1}^{(d)}(X,\theta)=\delta(\sin(\theta-\theta_o)),~~~~\theta_o \in [0,\pi];
\edm
\bdm
w_{\hat q}^{(d)}(X,\theta)=X\cos\theta\delta(\sin\theta);
\edm
\bdm
w_{\hat p}^{(d)}(X,\theta)=X\delta(\theta-\pi/2);
\edm
\bdm
w_{\hat q\hat p}^{(d)}(X,\theta)=X^2\delta(\theta-\pi/4)-\frac{1}{2}X^2\delta(\sin\theta)
-\frac{1}{2}X^2\delta (\theta-\pi/2)+\frac{i}{2\pi}.
\edm

Note, that for symplectic tomography the quantizer
and dequantizer operators are given by the formulas
\bdm
\hat U(X,\mu,\nu)=\delta(X\hat 1-\hat q\mu-\hat p\nu),
\edm
\bdm
\hat D(X,\mu,\nu)=\frac{1}{2\pi}
~e^{\ds{i(X-\hat q\mu-\hat p\nu)}},
\edm
and for the corresponding symbol and dual symbol for the operator $\hat A$
we have
\bdm
w_{\hat A}(X,\mu,\nu)=\mbox{Tr}\{\hat A\hat U(X,\mu,\nu) \},~~~
\hat A=\int w_{\hat A}(X,\mu,\nu)\hat D(X,\mu,\nu)
\mbox{d}X~\mbox{d}\mu~\mbox{d}\nu,
\edm
\bdm
w_{\hat A}^{(d)}(X,\mu,\nu)=\mbox{Tr}\{\hat A\hat D(X,\mu,\nu) \},~~~
\hat A=\int w_{\hat A}^{(d)}(X,\mu,\nu)\hat U(X,\mu,\nu)
\mbox{d}X~\mbox{d}\mu~\mbox{d}\nu.
\edm

\section{Representation of symbols of operators in terms of regular generalised functions}
The dual symbols of the operators for the optical probability allow
the representation in terms of regular generalised functions.

The dual symbol $w_{\hat A}^{(d)}(X,\theta)$
of some operator  $\hat A$
defines the linear continuous functional on the set of optical distributions
$w(X,\theta),$ belonging to space ${\cal S}^{2n}$ of well-behaved test functions.
Thus, the set of $w_{\hat A}^{(d)}(X,\theta)$ actually 
defines the set of generalised functions on ${\cal S}^{2n}.$
Obviously, that the equality of two symbols of one operator 
have to define as the functional equality or equality of two generalised functions,
i.e. two symbols are equal each other when for any distribution 
$w(X,\theta)\in{\cal S}^{2n}$ we have the equality of values of corresponding functionals,
denoted by these symbols.
Thus there are set of symbols for any operator $\hat A$ which are equal each other
in the meaning of generalized functions.

In the previous paragraph  we have found the general expression for the dual symbol of 
arbitrary operator and presented the singular forms of some operators.
Singular forms of operators are convinient for analitical calculations, but for the
numerical calculations and for the processing of experimental data the representation 
of symbols in the  form of regular generalised functions can be more preferable. 

If $\hat A$ is an arbitrary operator in Wigner representation
with  existing average value, then the integral
\bdm
\int{\bf\mbox R}[\hat A~W(\vec q,\vec p)](\vec X,\vec\theta)
{\mbox d}^nX=\langle\hat A\rangle
\edm
does not depend on $\vec\theta$.
Taking into account the definition of dual symbol of operator
we can write
\begin{eqnarray}
\langle\hat A\rangle&=&\int w_{\hat A}^{(d)}(\vec X,\vec\theta)
w(\vec X,\vec\theta){\mbox d}^nx{\mbox d}^n\theta\nonumber \\[3mm]
&=&
\frac{1}{\pi^n}\int{\bf\mbox R}[\hat AW(\vec q,\vec p)](\vec X,\vec\theta)
{\mbox d}^nX{\mbox d}^n\theta=
\frac{1}{\pi^n}\int{\bf\mbox R}[\hat A]
{\bf\mbox R}[W(\vec  q,\vec p)](\vec X,\vec \theta)
{\mbox d}^nX{\mbox d}^n\theta.
\label{r41_1}
\end{eqnarray}
where ${\bf\mbox R}[\hat A]$ is an explicit form of the operator $\hat A$ 
in the optical tomographic representation which we found in Sec. 3. 
In turn, the continuous linear functionals (generalized functions) of the form 
\bdm
\frac{1}{\pi^n}\int{\mbox d}^nX{\mbox d}^n\theta{\bf\mbox R}[\hat K]
w(\vec X,\vec \theta)
\edm
acting on the set of marginal distributions of $w(\vec X,\vec\theta),$ found
by the above rules easily can be represented in the form of regular
generalized functions. 
For example, let us regularize the functional 
$\int[\partial/\partial X_i]^{-2}w(\vec X,\vec\theta){\mbox d}^nX.$
After a double integration by parts we have: 
\begin{eqnarray}
\int\left[\frac{\partial}{\partial X_i}\right]^{-2}
w(\vec X,\vec\theta){\mbox d}^nX&=&
\int{\mbox d}^{n-1}X\left\{
X_i\left.\left[\frac{\partial}{\partial X_i}\right]^{-2}
w(\vec X,\vec\theta)\right|_{-\infty}^{+\infty}\right.\nonumber \\[3mm]
&-&\left.\left.
\frac{X_i^2}{2}\left[\frac{\partial}{\partial X_i}\right]^{-1}
w(\vec X,\vec\theta)\right|_{-\infty}^{+\infty}+
\int\limits_{-\infty}^{+\infty}\frac{X_i^2}{2}
w(\vec X,\vec\theta){\mbox d}X_i\right\}.
\label{r41_2}
\end{eqnarray}
Clarify the behavior of $x_i[\partial/\partial X_i]^{-2}w(\vec X,\vec\theta)$
when $x_i\to\pm\infty$
\be                             \label{r41_3}
X_i\left[\frac{\partial}{\partial X_i}\right]^{-2}w(\vec X,\vec\theta)=
X_i\int W(\vec X,\vec p)\frac{
\exp(-ik_\sigma(X_\sigma-q_\sigma\cos\theta_\sigma-p_\sigma\sin\theta_\sigma))}
{(-ik_i)^2}{\mbox d}^nk{\mbox d}^n q{\mbox d}^np.
\ee
We can see that the expression under the integral oscillates strongly
at  $k_i$ when $x_i\to\pm\infty$ and final $q_i$ и $p_i,$
but when $q_i,$ $p_i$ $\to\pm\infty$ we have $W({\bf q},{\bf p})\to0$ exponentially.
Thus whole the integral (\ref{r41_3}) $\to0$ when  $x_i\to\pm\infty.$
Similarly, we can find that the second term on the right side of the integrand
(\ref{r41_2}) is zero. Finally, we obtain 
\begin{eqnarray}
&&\int\left[\frac{\partial}{\partial X_i}\right]^{-2}
w(\vec X,\vec \theta){\mbox d}^nX=
\int\frac{X_i^2}{2}
w(\vec X,\vec\theta){\mbox d}^nX,~~~~{\mbox{ or}}\nonumber \\[3mm]
&&\left[\frac{\partial}{\partial x_i}\right]^{-2}\longrightarrow
\frac{x_i^2}{2}.\nonumber
\end{eqnarray}
We write down the regular representations of the operators most frequently
used in intermediate calculations: 
\be     \label{r42_2}
\begin{tabular}{lclllcl}
$\ds{\frac{\partial}{\partial X_i}}$&$\to$&$0;$&~~~~&
$X_i\ds{\frac{\partial}{\partial X_i}}$&$\to$&$-1;$\\[3mm]

$\ds{\left[\frac{\partial}{\partial x_i}\right]^{-1}}$&$\to$&$-X_i;$&~~~~&
$X_i\ds{\left[\frac{\partial}{\partial X_i}\right]^{-1}}$&$\to$&$\ds{-\frac{X_i^2}{2};}$\\[3mm]

$\ds{\frac{\partial^2}{\partial\theta_i^2}}$&$\to$&
\multicolumn{5}{l}{
$-\delta'(\theta_i-\pi)+\delta'(\theta_i)=0;$
}
\end{tabular}
\ee

\be     \label{r42_3}
\begin{tabular}{lclllcl}
$\sin\theta_i\ds{\frac{\partial}{\partial\theta_i}}$&$\to$&$-\cos\theta_i;$&~~~~&
$\sin^2\theta_i\ds{\frac{\partial}{\partial\theta_i}}$&$\to$&$-\sin2\theta_i;$\\[3mm]

$\ds{\frac{\sin2\theta_i}{2}\frac{\partial}{\partial\theta_i}}$&$\to$&$-\cos\theta_i;$&&
$\sin^2\theta_i\ds{\frac{\partial^2}{\partial\theta_i^2}}$&$\to$&$2\cos2\theta_i;$\\[3mm]

$\ds{\frac{\sin2\theta_i}{2}\frac{\partial^2}{\partial\theta_i^2}}$&$\to$&
\multicolumn{5}{l}{
$-(\delta(\theta_i-\pi)-\delta(\theta_i))-2\sin2\theta_i=-2\sin2\theta_i;$
}\\[3mm]

$\cos\theta_i\ds{\frac{\partial}{\partial\theta_i}}$&$\to$&
\multicolumn{5}{l}{
$-(\delta(\theta_i-\pi)+\delta(\theta_i))+\sin\theta_i$
}
\end{tabular}
\ee
The presence of $\delta-$ functions and their derivatives in these formulas
is a removable singularity due to the fact that the final regularization occurs
on the set of functions $w(\vec X,\vec\theta),$ with the additional symmetry
property
\bdm
\frac{\partial^k}{\partial\theta^k}w(X_i,\theta_i=0)=
\frac{\partial^k}{\partial\theta^k}w(-X_i,\theta_i=\pi),~~{\mbox{where}}
~~k=0,1,2,3\dots,
\edm
In addition, we give one more useful formula for the dual
symbol of the product of two operators. 
Let the operators $\hat{\cal A}$ and $\hat{\cal B},$ act on the set of such
$W(\vec q,\vec p)\in{\cal S}^{2n},$ that for every
$W(\vec q,\vec p)\in{\cal S}^{2n}$ the function
$\hat{\cal A}W(\vec q,\vec p)$ and $\hat{\cal B}W(\vec q,\vec p)$
also belong to ${\cal S}^{2n},$ then from (\ref{r38_1})  for the product
of two operators and from the formula (\ref{r41_1}) we have equality: 
\bdm
\langle\hat{\cal A}\hat{\cal B}\rangle=\int w_{\hat{\cal A}\hat{\cal B}}^{(d)}
(\vec X,\vec\theta)w(\vec X,\vec\theta){\mbox d}^nX{\mbox d}^n\theta
=\int w_{\hat{\cal A}}^{(d)}(\vec X,\vec\theta)
{\bf\mbox R}[\hat{\cal B}]{\bf\mbox R}[W(\vec q,\vec p)](\vec X,\vec\theta)
{\mbox d}^nX{\mbox d}^n\theta,
\edm
or
\bdm
w_{\hat{\cal A}}^{(d)}(\vec X,\vec\theta)
{\bf\mbox R}[\hat{\cal B}](\vec X,\vec\theta)\longrightarrow
w_{\hat{\cal A}\hat{\cal B}}^{(d)}
(\vec X,\vec\theta).
\edm
Let us find a regular symbol of the operator $q_i$.
From  (\ref{r46_1}), (\ref{r41_1}),  (\ref{r42_2}) and (\ref{r42_3})
we have the chain of equalities: 
\begin{eqnarray}
\langle\hat q_i\rangle&=&\int f_{\hat q_i}~w~
{\mbox d}^nX{\mbox d}^n\theta=\frac{1}{\pi^n}\int{\bf\mbox R}[\hat q_iW]
{\mbox d}^nX{\mbox d}^n\theta\nonumber \\[3mm]
&=&\frac{1}{\pi^n}\int\left\{\sin\theta_i
\left[\frac{\partial}{\partial X_i}\right]^{-1}\frac{\partial}{\partial\theta_i}+
X_i\cos\theta_i\right\}w~{\mbox d}^nX{\mbox d}^n\theta=
\frac{2}{\pi^n}\int X_i\cos\theta_i~w~{\mbox d}^nX{\mbox d}^n\theta.\nonumber
\end{eqnarray}
Thus we can write
\bdm
w_{\hat q_i}^{(d)}(\vec X,\vec\theta)=\frac{2}{\pi^n}X_i\cos\theta_i.
\edm
Similarly, we find
\bdm
\begin{tabular}{lll}
$w_{\hat p_i}^{(d)}(\vec X,\vec\theta)=\ds{\frac{2}{\pi^n}X_i\sin\theta_i;}$&~~~~&
$w_{\hat q_i^2}^{(d)}(\vec X,\vec\theta)=\ds{\frac{X_i^2}{\pi^n}(1+2\cos2\theta_i);}$\\[3mm]
$w_{\hat p_i^2}^{(d)}(\vec X,\vec\theta)=\ds{\frac{X_i^2}{\pi^n}(1-2\cos2\theta_i);}$&~~~~&
$w_{\hat q_i\hat p_i}^{(d)}(\vec X,\vec\theta)=\ds{\frac{2}{\pi^n}X_i^2\sin2\theta_i
+\frac{i}{2\pi^n}.}$
\end{tabular}
\edm
From (\ref{r46_5}) we can find symbols of the components of the angular
momentum of the particle 
\bdm
w_{\hat l_1}^{(d)}(X_1,X_2,X_3;\theta_1,\theta_2,\theta_3)=\frac{4}{\pi^{2n}}
X_2X_3\sin(\theta_3-\theta_2);
\edm
\bdm
w_{\hat l_2}^{(d)}(X_1,X_2,X_3;\theta_1,\theta_2,\theta_3)=\frac{4}{\pi^{2n}}
X_3X_1\sin(\theta_1-\theta_3);
\edm
\bdm
w_{\hat l_3}^{(d)}(X_1,X_2,X_3;\theta_1,\theta_2,\theta_3)=\frac{4}{\pi^{2n}}
X_1X_2\sin(\theta_2-\theta_1);
\edm
From formulas (\ref{r34_4}), (\ref{r38_3}),  (\ref{r42_2}) and (\ref{r42_3}) and we can find
\bdm
w_{\hat a_i}^{(d)}(\vec X,\vec\theta)=\ds{\frac{\sqrt2}{\pi^n}}
X_i(\cos\theta_i+i\sin\theta_i);~~~~
w_{\hat a_i^{\dag}}^{(d)}(\vec X,\vec\theta)=\ds{\frac{\sqrt2}{\pi^n}}
X_i(\cos\theta_i-i\sin\theta_i);
\edm
\bdm
w_{\hat a_i^{\dag}\hat a_i}^{(d)}(\vec X,\vec\theta)=
w_{\hat N_i}^{(d)}(\vec X,\vec\theta)={\ds\frac{1}{\pi^n}}
(X_i^2-1/2).
\edm

Having different forms of dual symbols of the same operators, we can write test expressions
for the experimentally measured tomograms. 
Thus, for the $\hat q$ quadrature we have
\be		\label{test1q}
\langle\hat q\rangle=\int X\cos\theta\delta(\sin\theta)
w(X,\theta) \mbox{d}X~\mbox{d}\theta 
=\frac{2}{\pi}\int X\cos\theta w(X,\theta)
\mbox{d}X~\mbox{d}\theta. 
\ee
Similar test expressions can be writen for the other operators. 

\section{Conclusion}
\pst
To summarize, we point out the main results of this work.

We obtained the correspondence rules and explicit expressions for 
operators of physical quantities in optical tomographic representation.
We presented the general formalism for symbols of operators in this representation.
We found an explicit expressions for dual symbols of physical
quantities in terms of regular generalised functions.
The expressions for operators  found in the
work provide the possibility of direct calculations of physical quantities from the optical tomogram
without recalculation of it to the Wigner function or to the density matrix.

\section*{Acknowledgments}
V.I.M. thank the Russian Foundation for Basic
Research for partial support under Projects Nos. 09-02-00142 and 10-02-00312.

\newpage

\end{document}